\documentclass[onecolumn,prx,superscriptaddress,bibnotes,notitlepage,nofootinbib, floatfix]{revtex4-1}
\usepackage{color}
\usepackage[normalem]{ulem}
\usepackage{makecell}
\usepackage{tabularx}
\definecolor{red}{rgb}{0.75,0,0}
\definecolor{blue}{rgb}{0,0,0.75}
\definecolor{green}{rgb}{0,0.5,0}

\usepackage{mathrsfs}
\usepackage{amsmath}
\usepackage{amssymb}
\usepackage{lipsum}
\usepackage{bm}
\usepackage{xcolor}
\usepackage{graphicx}
\usepackage{verbatim}
\usepackage[T1]{fontenc}
\usepackage{hyperref}
\usepackage{xcite}
\usepackage{zref-xr}
\zxrsetup{tozreflabel=false, toltxlabel=true, verbose}
\zexternaldocument*{film_nem_ch_Supp_arxiv}
\externalcitedocument{film_nem_ch_Supp_arxiv}

\def\be{\begin{equation}}
\def\ee{\end{equation}}
\def\bea{\begin{eqnarray}}
\def\eea{\end{eqnarray}}

\def\besub{\begin{subequations}}
\def\eesub{\end{subequations}}

\def\bwd{\begin{widetext}}
\def\ewd{\end{widetext}}

\definecolor{ao(english)}{rgb}{0.0, 0.5, 0.0}
\definecolor{armygreen}{rgb}{0.29, 0.33, 0.13}
\definecolor{auburn}{rgb}{0.43, 0.21, 0.1}
\definecolor{brightmaroon}{rgb}{0.76, 0.13, 0.28}
\definecolor{cadmiumred}{rgb}{0.89, 0.0, 0.13}
\definecolor{carnelian}{rgb}{0.7, 0.11, 0.11}
\definecolor{cornellred}{rgb}{0.7, 0.11, 0.11}
\definecolor{crimsonglory}{rgb}{0.75, 0.0, 0.2}
\definecolor{orangeyellow}{rgb}{0.3, 0.2, 0.2}
\definecolor{fluorescentorange}{rgb}{1.0, 0.75, 0.0}
\definecolor{gamboge}{rgb}{0.89, 0.61, 0.06}
\newcommand{\bsf}[1]{\textsf{\textbf{#1}}}

\newcommand{\AM}[1]{\textcolor{black}{#1}}
\newcommand{\AMF}[1]{\textcolor{black}{#1}}

\begin{document}
\title{Floating flocks: Two-dimensional long-range uniaxial order in three-dimensional active fluids }
\author{Ananyo Maitra}
\email{nyomaitra07@gmail.com}
\affiliation{Laboratoire de Physique Th\'eorique et Mod\'elisation, CNRS UMR 8089,
	CY Cergy Paris Universit\'e, F-95302 Cergy-Pontoise Cedex, France}
\affiliation{Sorbonne Universit\'{e} and CNRS, Laboratoire Jean Perrin, F-75005, Paris, France}

\begin{abstract}
\AMF{Elongated active units cannot spontaneously break rotation symmetry in bulk fluids to form nematic or polar phases. This has led to the image of active suspensions as spontaneously evolving, spatiotemporally chaotic fluids. In contrast, I show that bulk active fluids have stable active nematic and polar states at fluid-fluid or fluid-air interfaces. The active flow-mediated long-range interactions that destroy the ordered phase in bulk, lead to long-range order at the interface. The active fluids have a surface ordering transition and form states with quiescent, ordered surfaces and a chaotic bulk. I further consider active units that are constrained to live at an interface to examine the minimal conditions for the existence of two-dimensional order in bulk three-dimensional fluids. In this case, immotile units do not order, but motile particles still form a long-range-ordered polar phase. This prediction of stable, uniaxial, active phases in bulk fluids may have functional consequences for active transport.}
\end{abstract}

\maketitle

\section{Introduction}
Active matter theories describe systems whose constituents convert a sustained supply of energy into work  \cite{SRJSTAT, LPDJSTAT, RMP, Prost_nat, Sal1, SRrev}. The microscopic energy input leads to macroscopic forces and currents, dramatically modifying the mechanics and statistical mechanics of active systems relative to their dead or \AMF{equilibrium} counterparts. This has spectacular effects on the phases and phase behaviours that active systems display. 

One of the most notable consequences of the microscopic drive on the phase behaviour of active systems is the threshold-free instability of uniaxial phases {-- both nematic and polar --} in incompressible, bulk Stokesian fluids \cite{Aditi1, Voit1, RMP} first discussed by Simha and Ramaswamy \cite{Aditi1}. This bulk Simha-Ramaswamy instability -- via which an oriented active fluid is driven to a spatiotemporally chaotic state \cite{Alert1, Alert_Rev} -- has become the defining image of active suspensions. The impossibility of a bulk uniaxial state implies that active suspensions do not \AM{spontaneously} break rotation symmetry i.e., there is no bulk isotropic-nematic or isotropic-polar transition in Stokesian fluids. 
Instead, activity constantly stirs the bulk Stokesian fluid via a random stress with a correlation length vanishing as the inverse of the square root of activity. This has led to the expectation that uniaxial phases \AM{cannot} exist in momentum-conserved active \AM{systems, in contrast to fluids on substrates where such order can not only exist at an arbitrarily high active drive, but can also be strengthened by it \cite{Ano_apol, Ano_pol, Niladri1, Niladri2}.}

In this article, I show that while \emph{bulk} uniaxial ordering is indeed forbidden in active suspensions, two-dimensional nematic and polar \AM{phases that \emph{spontaneously} break a \emph{continuous} symmetry} exist at interfaces or boundaries of momentum-conserved fluids. That is, a bulk suspension of elongated active units in a fluid can have stable, long-range ordered nematic or polar wetting layers at fluid interfaces or boundaries. 
 Interface-associated \AM{ordering transitions and} phases have been examined in detail in \AMF{equilibrium} systems \cite{int_ord_nem, Liu_Toner, Hohenberg, Lubensky1, Lubensky2}. However, in \AMF{equilibrium} magnets \cite{Hohenberg, Lubensky1, Lubensky2} and nematic fluids \cite{Poniewerski1, int_ord_nem}, while such transitions precede the bulk transition, as the bulk critical point is approached from above, the thickness of the interfacial ordered phase diverges. In contrast, since active fluids always remain disordered in the bulk, the thickness of ordered wetting layers does not change appreciably with noise strength.
  
 Furthermore, in \AMF{equilibrium} \AM{materials}, interface-associated phase\AM{s} \AM{are more susceptible to random fluctuations than} bulk \AM{order}. In three-dimensional matter in equilibrium, a two-dimensional boundary-associated phase displays only quasi-long-range order \cite{int_ord_nem}, while the bulk has true long-range order since \AM{the effect of} Goldstone fluctuations \AM{is} more dominant in two dimensions than in three. In contrast, I show that both \AM{active} nematic and polar phases at interfaces or boundaries of bulk fluids display true \emph{long-range} order (LRO). 
This \emph{stabilisation} of interfacial order in momentum-conserved systems is due to active fluid flows; while such flows \emph{destabilise} bulk uniaxial order, they \emph{anomalously stabilise} interfacial order. In other words, while active flows ``anti-screen'' fluctuations of bulk orientational order, they \emph{screen} fluctuations of \emph{interfacial} uniaxial states in the \emph{same system}. This is akin to observing a Jeans-like instability in the bulk but a Coulomb-like screening at the interface. 
In bulk fluids, the ratio of active stress and viscosity introduces an inverse timescale \cite{SRrev}. However, this need not be proportional to the growth rate of an instability; it could be proportional to the \emph{relaxation rate} of a massive mode. It turns out to be proportional to the \emph{growth rate} of the instability of an \AM{aligned} phase in bulk fluids \AM{ultimately} because \AM{both total mass and total momentum are conserved}.
The crucial distinction that allows active flows to stabilise boundary-associated \AM{nematic} order while destabilising bulk order is that \AM{neither the number of active units nor the total mass at the interface is conserved. That is, both active units and the fluid can escape in the third dimension, with the two-dimensional interfacial flow-field not being incompressible even though the bulk three-dimensional flow is \cite{Liu_Toner, Prasad, com2}}. \AM{Interfacial \emph{polar} order is more robust.}

Beyond the fundamental physical interest of realising boundary-associated ordered phases in momentum-conserved active fluids, when no corresponding bulk phase exists, \AM{the fact that orientational order \AMF{may exist} \emph{even} in highly active fluids has}
important experimental and biological consequences. Some of the more widely used experimental systems for studying pattern formation in active uniaxial systems are composed of motor-microtubule filaments either at a two-fluid interface \cite{Dogic1, Sagues} or forming a self-assembled layer immersed in a bulk fluid \cite{Ano_und, Ramin_und}. While it had been assumed that uniaxial phases are forbidden in these geometries \cite{Guo1, Guo2}, this article demonstrates that this conclusion is contingent on experimental details.
Swimmers, such as bacteria \cite{bac_int1, bac_int2, bac_int3, bac_int4}, generally aggregate at interfaces \cite{com2} and may form effectively two-dimensional uniaxial phases. In cellular systems, uniaxial ordering is likely to be associated with interfaces or membranes -- such as in the cortex -- and is relevant for active transport \cite{Madan1, Madan2, jitu1, jitu2}. Finally, there has recently been a great deal of interest in the possibility of forming protocells by microphase separation in active fluids. If such protocells contain elongated particles -- for instance, if the phase separation leads to droplets that are rich in elongated active units in a background that is poor in them -- they can form an ordered boundary layer of orientable filaments in the droplet or a protocortex.

I now \AM{describe the key question and results of this paper.}

\subsection{Summary of results}
\begin{table}[!htb]
	\begin{ruledtabular}
		\begin{tabular}{p{1.4in} p{1.4in}p{0.05in}p{1.4in}p{0.05in}p{1.4in}}
			& Diffusing in the bulk & & Living at the interface & & Incompressible\\
			\hline
			Nematic & LRO  & & Unstable  & & Unstable \\ 
			
			Polar & LRO  & & LRO  with GNF & & Unstable 
		\end{tabular}
	\end{ruledtabular}
	\caption{Summary of active uniaxial interfacial order depending on whether active units can be exchanged between the interface and the bulk or are confined to the interface or the interface is incompressible. LRO implies that a long-range ordered phase exists for some parameter regime in that category while GNF signifies giant number fluctuations.}
	\label{restable}
\end{table}
\AM{I show that active uniaxial particles in momentum-conserved fluids spontaneously break rotation symmetry and form long-range ordered polar or nematic phases at fluid-fluid or fluid-air interfaces. The hydrodynamic properties of these LRO uniaxial phases belong to a new universality class that I exactly characterise.
How do these interfacial ordered phases escape the Simha-Ramaswamy instability that doesn't allow bulk ordering? Is it simply because the two-dimensional velocity field at the fluid interface is compressible \cite{Liu_Toner, Prasad, com2}? To examine this, I consider active particles that are confined \emph{only} to a fluid interface and \emph{cannot} diffuse freely into the bulk fluid. The two-dimensional interfacial velocity is not \AMF{divergence-free} in this system either but, importantly, the number of active particles at the interface is conserved. This reveals that the minimal conditions for the existence of interfacial order differ for polar and apolar systems -- while a homogeneous nematic phase is \emph{generically} unstable in this case, the polar phase is more robust. The form of the instability is however distinct from that in incompressible bulk fluids, and the lengthscale of the fluid structures formed at large active drive becomes independent of activity. I show that this is analogous to a previously unreported instability of aligned states in bulk, \emph{compressible}, momentum-conserved active fluids. Thus, active nematic order at fluid interfaces is only stable when neither the number of active units nor the total mass at the interface is conserved. Motile particles escape this instability and form an LRO state for large enough propulsion speeds.} The concentration fluctuations in this motile phase display giant number fluctuations (GNF) with the R.M.S. number fluctuations in a region containing on average  $\langle N\rangle$ swimmers scaling as $\langle N\rangle^{3/4}$ instead of $\sqrt{\langle N\rangle}$ as it would in all equilibrium systems. \AM{The results for the different geometries are summarised in Table \ref{restable}. 
}

\section{Model description}
\label{mod_desc}
\begin{figure}[!htb]
	\centering
	\includegraphics[width=16cm]{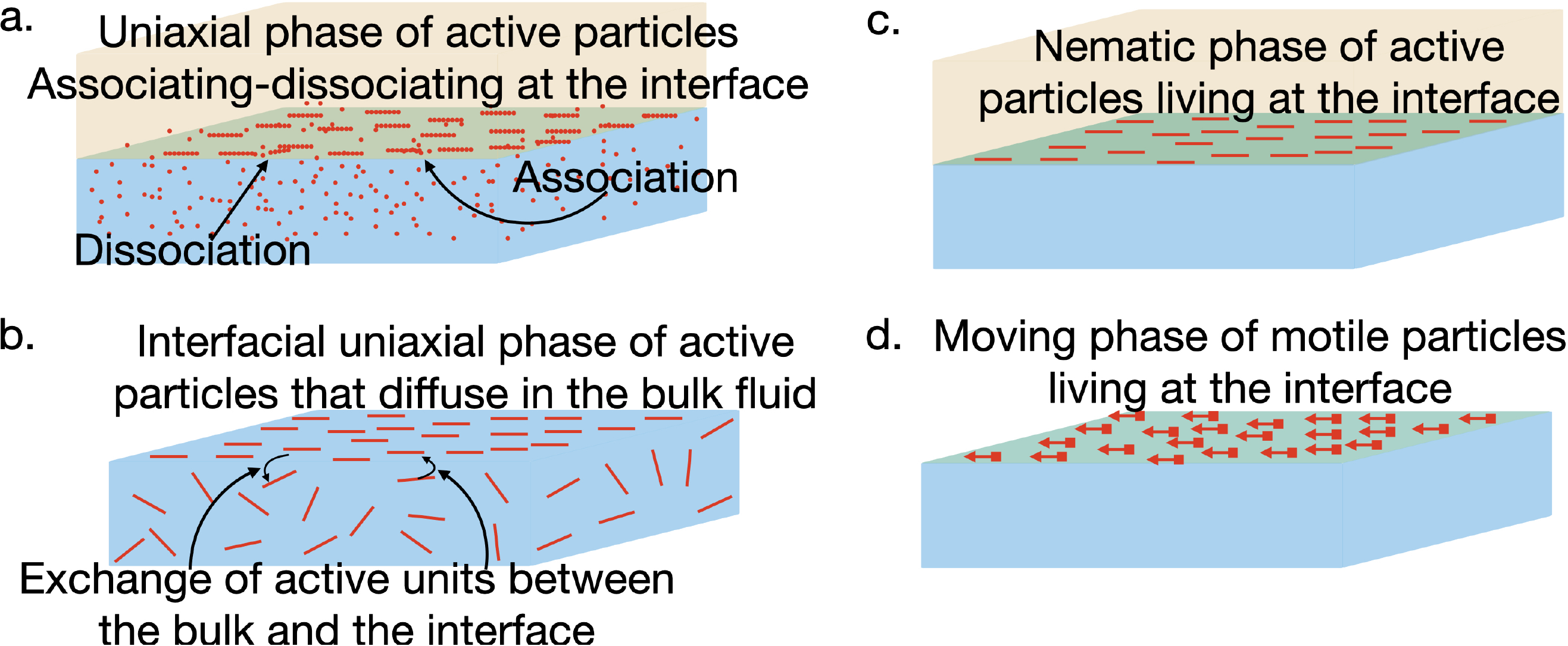}
	\caption{a. and b. displays two experimental situations described by the NNC model): a. A nematic phase formed by polymeric elongated active particles which associate and dissociate at the interface between two fluids, and b. An apolar phase formed by elongated active units at the boundary of a bulk fluid where the active units are exchanged between the bulk and the interface and diffuse in the bulk fluid. c. A schematic for a system described by NC model: a possible nematic phase formed by elongated, active particles living at the interface between two fluids. The model also describes a system of active rods living at the boundary of a bulk fluid. d. A polar phase of motile particles confined at the surface of a bulk fluid. The hydrodynamics of this phase is described by the PC model. Equivalently, the model also describes the dynamics of an actively moving phase of polar particles confined at a two-fluid interface.}
	\label{Fig_diss}
\end{figure}
\AM{In this article, I consider the ordering of elongated active units at fluid-fluid or fluid-air interfaces. The flat interface is assumed to be clean i.e., not coated by surfactants, and is taken to be situated at $z=0$. \AMF{While the three-dimensional rotation symmetry of the bulk fluid is broken at $z=0$ by the interface, the interfacial plane has $\mathbb{U}(1)$ symmetry i.e., two-dimensional rotation symmetry. The elongated active units at the interface are taken to be preferentially predisposed to lie parallel to it, in the models I consider. That is, any out-of-plane component of the orientation of active units at the interface is taken to relax to zero, fast. This may be due to planar anchoring conditions at the interface or active torques \cite{Juho, Ardekani}. 
		However, since the in-plane axes \cite{int_ord_nem} are degenerate without any preferred direction,
		to form aligned states at the interfacial plane, elongated active units must \emph{spontaneously} break $\mathbb{U}(1)$ invariance. I consider such spontaneous breaking of two-dimensional rotation symmetry at fluid interfaces. 
}}

\AM{\subsection{Interfaces in bulk active fluids}}
\AM{In the first class, I consider bulk fluids containing either active orientable units or components that form such objects. In these cases, the number of active units at the interface is \emph{not} conserved since active units diffuse into the bulk. \AM{I consider both nematic and polar order, which will be accounted for via the NNC (nematic, non-conserved) and PNC (polar, non-conserved) models.}
	The same hydrodynamic models apply to two distinct physical situations: i. when the active units themselves diffuse in the bulk fluid and ii. when monomers associate to form active units at the interface as well as dissociate and diffuse in the bulk. These two situations are depicted in Fig. \ref{Fig_diss} a and b (respectively at a fluid-air and a two-fluid interface).}
	The first models systems in which elongated active swimmers, such as bacteria, form an ordered phase at an interface but can move into the bulk fluid. The second is relevant when monomers associate or dissociate at a fluid-fluid or fluid-air interface to form polymers which then become active due to the action of motors. This is the case, for instance, in the cellular cortex \cite{Sal1, Salbreaux}. 
	
	\AM{In systems described by this class of models, the number of active units at the interface is not conserved even though their \emph{total} number (or the total number of monomers that compose the active units), both in the bulk and at the interface, is conserved. The dynamical equations of the in-plane order parameters only directly couple with the in-plane density of active particles (and the in-plane velocity). Because an excess in-plane concentration of active particles does not relax by diffusing through the interfacial layer in this case but by diffusing into the bulk fluid, the interfacial concentration of active units is not a slow variable -- it relaxes within a microscopic time to a value determined by the bulk concentration of active units. The bulk fluid acts as a bath of particles that holds the interfacial particle concentration locally fixed \cite{Malthus, Khandkar}. Thus, the interfacial concentration is not a hydrodynamic variable in NNC and PNC models. While the total concentration of active units (or monomers) in the bulk and the interface \emph{is} a hydrodynamic variable, because of the incompressibility of the bulk fluid, this doesn't affect the in-plane interfacial ordering. I show this explicitly in Supp. \ref{App25}. I now construct the dynamical equations for NNC and PNC models.}

	\paragraph{NNC Model: Interfacial active nematic composed of nematogens diffusing in the bulk}
	\label{INdesc}
	\AM{Since in both NNC and PNC models, the interfacial concentration of active units is not a hydrodynamic variable, the long-time, large-scale evolution of an in-plane ordered state is determined by the coupled dynamics for an in-plane order parameter that measures the degree of interfacial ordering of filaments and the in-plane velocity field. In nematic systems, the relevant order parameter is a rank-2 tensor}
	\begin{equation}
		{\bsf Q}=\frac{S}{2}\begin{pmatrix}\cos 2\theta & \sin 2\theta\\ \sin 2\theta & -\cos 2\theta\end{pmatrix}
	\end{equation}
	where I take the ordering direction to be along $\hat{x}$, $\theta$ is the deviation of the local nematic order from $\hat{x}$ and $S$ is the magnitude of the nematic order whose steady state value is $S_0=\langle S\rangle$. 
	\AM{This couples to the two-dimensional, interfacial flow field ${\bf v}$ as}
	\begin{equation}
		\label{QtendynMT}
		\dot{\bsf Q}={\bsf Q}\cdot\boldsymbol{\Omega}-\boldsymbol{\Omega}\cdot{\bsf Q}-{\lambda}{\bsf A}^{ST}-\Gamma{\bsf {H}} +\boldsymbol{\xi}^Q,
	\end{equation}
	where the overdot denotes the convected derivative $\partial_t+{\bf v}\cdot\nabla_\perp$, \AM{$\nabla_\perp\equiv (\partial_x,\partial_y)$,} $\boldsymbol{\Omega}=(1/2)[\nabla_\perp{\bf v}-(\nabla_\perp{\bf v})^T]$ is the vorticity tensor at the interface, ${\bsf A}=(1/2)[\nabla_\perp{\bf v}+(\nabla_\perp{\bf v})^T]$ is the strain-rate tensor at the interface, with $|\lambda|>1$ describing nematogens with tendency to align under an imposed shear flow and $|\lambda|<1$ describing flow tumbling, the superscript $ST$ denotes symmetrized, traceless part of a tensor, ${\bsf H}=[\delta F_{Q}/\delta{\bsf Q}]^{ST}$ and $\boldsymbol{\xi}^Q$ is a non-conserving, \AM{Gaussian, white} noise with the \AM{standard deviation $2\Delta^Q$}. 
	Finally, $\Gamma$ controls the relaxation to the equilibrium steady state in the absence of \AM{flow}.
	The standard Landau-de Gennes free energy for a two-dimensional nematic, written here in a simplified one Frank constant approximation, is $F_{Q}=\int d{\bf r}_\perp f_Q$, with 
	$f_Q=({\alpha}/{2})({\bsf Q}:{\bsf Q})+({\beta}/{2})({\bsf Q}:{\bsf Q})^2+({K}/{2})(\nabla_\perp{\bsf Q})^2$,
	which supports orientational order for $\alpha<0$. \AM{Note that the dynamical equation for the nematic order parameter at this order in gradients and fields is equivalent to that of an \AMF{equilibrium} system. There is no active term in this equation because any such term involving purely ${\bsf Q}$ can be absorbed by redefining $f_Q$. The lowest order non-integrable term has two powers of gradient and two powers of ${\bsf Q}$ and, as I will show post-facto, is irrelevant. Further, note that the entry of active particles into the interfacial layer at \emph{arbitrary} in-plane angles, and their exit from it, imply a local, stochastic change of the orientation field leading to an additional source of noise \cite{Khandkar, Malthus} (i.e., leads to a modification of $\Delta^Q$).}
	
	\AM{The in-plane velocity field ${\bf v}$ is obtained from the three-dimensional bulk velocity field ${\bf V}$ as ${\bf V}_\perp|_{z=0}={\bf v}$, where $\perp$ represents the $x$ and $y$ components of a vector. For slow flows, ${\bf V}$ satisfies the incompressible Stokes equation forced by a surface force density ${\bf f}^s$ which, in momentum-conserved systems, can be written as a divergence of an in-plane stress $\nabla_\perp\cdot\bm{\sigma}^s$. The hydrodynamic behaviour of the in-plane ordered state is controlled by an interfacial active stress $\bm{\sigma}^s=-\zeta{\bsf Q}$ where $\zeta>0$ denotes an extensile suspension while one with $\zeta<0$ signifies contractility \cite{Cates_Maren} (see Supp. \ref{App251} for a discussion of \AMF{detailed-balance obeying} subdominant terms in $\bm{\sigma}^s$). The equations for ${\bsf Q}$ and ${\bf V}$ have to be solved simultaneously to examine the stability of interfacial order. I show in Supp. \ref{App2} that ${\bf v}$ appearing in \eqref{QtendynMT} is obtained by solving the Stokes equation for ${\bf V}$. It is related to ${\bf f}^s$ via ${\bf v}={\bsf M}\cdot{\bf f}^s$, where the mobility written in Fourier space, to the lowest order in wavenumbers, is}
		\begin{equation}
		\label{mobil_expMT}
		{\bsf M}=\frac{1}{4\eta|q_\perp|^3}\begin{pmatrix}q_x^2+2q_y^2 & -q_xq_y\\ -q_xq_y & 2q_x^2+q_y^2\end{pmatrix}
	\end{equation}
	with ${\bf q}_\perp\equiv(q_x, q_y)$ being the in-plane wavevector of a perturbation, and $\eta$ being the arithmetic mean of the viscosities of the fluid above and below the interface. \AM{See Supp. \ref{App2} for higher order in $q_\perp$ corrections to this, arising from a finite Saffmann-Delb\"{u}ck length.}
	
	\AM{Note that even though active units are present in the bulk fluid, since they do not order \cite{Aditi1}, bulk active forcing leads only to a conserving noise, correlated over finite spatial and temporal scales, \cite{Alert_Rev, Graham} which doesn't affect the hydrodynamic behaviour of the {interfacial ordered phase} (see Supp. \ref{App1} for a demonstration). }

	\paragraph{PNC Model: Interfacial active polar fluid composed of motile particles that diffuse in the bulk}
	\label{IPdesc}
	\AM{The dynamics of an interfacial polar state is described by the coupled dynamics of a polar order parameter ${\bf p}=p(\cos\theta, \sin\theta)$ -- here $p$ is the magnitude of the polar order and $\theta$ is the local deviation of the polarisation from the ordering direction, which is taken to be $\hat{x}$ -- and the in-plane interfacial velocity field ${\bf v}$. Just as in the NNC model, the interfacial concentration of polar active units is not a hydrodynamic variable. The dynamics of ${\bf p}$ is }
	\begin{equation}
		\label{polgendynMT}
		\dot{{\bf p}}+v_p{\bf p}\cdot\nabla{\bf p}+\boldsymbol{\Omega}\cdot{\bf p}=-{\lambda}{\bf p}\cdot{\bsf A}-\lambda_p\nabla_\perp^2{\bf v}-\Gamma_p{\bf h}+\boldsymbol{\xi}^p
	\end{equation}
	where $v_p$ denotes active self-advection due to the motility of the polar particles, $F=\int d{\bf r}_\perp [(\alpha/2){\bf p}^2+(\beta/4){\bf p}\AM{^4}+(K/2)(\nabla_\perp{\bf p}^2)]=\int d{\bf r}_\perp f_p$ is the standard free energy for polar liquid crystals, ${\bf h}=\delta F/\delta{\bf p}$ and $\boldsymbol{\xi}^p$ is a spatiotemporally white noise with a variance $2\Delta^p$. Other nonlinear gradient terms do not affect the hydrodynamics of the interfacial phase (see Supp. \ref{App251}) and have been suppressed here. \AM{As in the NNC model, the interfacial velocity is obtained by solving the interfacially-forced bulk Stokes equation as ${\bf v}={\bsf M}\cdot{\bf f}^s$. Here, ${\bsf M}$ is given by \eqref{mobil_expMT} and} ${\bf f}^s= -{\zeta}\nabla_\perp\cdot\left({\bf pp}-{{p^2\bsf I}}/{2}\right)$, which is equivalent to the active force in the NNC model.

\AM{\subsection{Active interfacial layer in fluids}}
\AM{In the second class of models that I consider, active units are \emph{constrained} to live at fluid-fluid or fluid-air interfaces. This is particularly relevant for experiments on motor-microtubule layers at two-fluid interfaces \cite{Dogic1, Sagues} in which the numbers of microtubule filaments and motors are conserved. I will use these models to better understand the conditions that are required for stable active interfacial order. Since the active units are constrained to live at the interface, their number at the interface \emph{is} conserved, unlike in the NNC and PNC models and their concentration $c$ is an additional hydrodynamic variable that couples to the in-plane order parameter and velocity fields. As earlier, I consider both nematic and polar order, accounted for by the NC (nematic, conserved) and PC (polar, conserved models), which I now describe.}
\paragraph{NC Model: Interfacial active nematic composed of nematogenic species living at the interface}
\label{IINdesc}
The dynamics of the ${\bsf Q}$ tensor \AM{in this model} is still described by \eqref{QtendynMT}. \AM{The interfacial velocity field is again obtained using the mobility in \eqref{mobil_expMT} and an interfacial 
stress }$\bm{\sigma}^s=-\zeta(c) {\bsf Q}-\Pi_c(c){\bsf I}$ where $\Pi_c$ is an isotropic pressure-like term, with both active and \AMF{non-active} contributions. Here, ${\bsf I}$ is the rank two identity tensor. The equation for $c$, to the lowest order in gradients, is simply $\partial_tc=-\nabla_\perp\cdot(c{\bf v})$. All other symmetry-allowed terms are subdominant to the ones noted here (see Supp. \ref{App251}).

\paragraph{PC Model: Interfacial active polar fluid composed of polar species living at the interface}
\label{IIPdesc}
The dynamics of the polarisation vector in the PC model is described by \eqref{polgendynMT} but with an additional free energy coupling between ${\bf p}$ and $c$: $F=\int d{\bf r}_\perp[f_p+\gamma{\bf p}\cdot\nabla_\perp c]$. The active stress \AM{-- the in-plane divergence of which gives the interfacial active force ${\bf f}^s$ --} is also concentration-dependant: $\bm{\sigma}^s=-{\zeta}(c)\left({\bf pp}-{{p^2\bsf I}}/{2}\right)-\Pi_c(c){\bsf I}$. Finally, \AM{the dynamical equation for the concentration field is $\partial_t c=-\nabla_\perp\cdot[c{\bf v}-v_c(c){\bf p}]$ where the first term accounts for the advection by the interfacial flow and $v_c(c)$ is the active motility.}
No other symmetry-allowed term affects the hydrodynamic theory of the interfacial polar phase \ref{App251}.

\section{Long-range ordered nematic and polar phase in NNC and PNC models}
\label{stabNNC}
I now demonstrate the central result of this article: bulk active fluids support \emph{stable}, {long-range} ordered active nematic and polar phases at (fluid-fluid or fluid-air) interfaces.
Because the bulk active fluid remains in the isotropic phase at all noise strengths \cite{Aditi1, Alert_Rev}, the interfacial ordered wetting layer never acquires a macroscopic thickness and there is no ordinary or extraordinary transition to a bulk-ordered phase \cite{DelGamma1, DelGamma2, Poniewerski1, int_ord_nem, Nakanishi, Sheng}.

\begin{figure}[!htb]
	\centering
	\includegraphics[width=8cm]{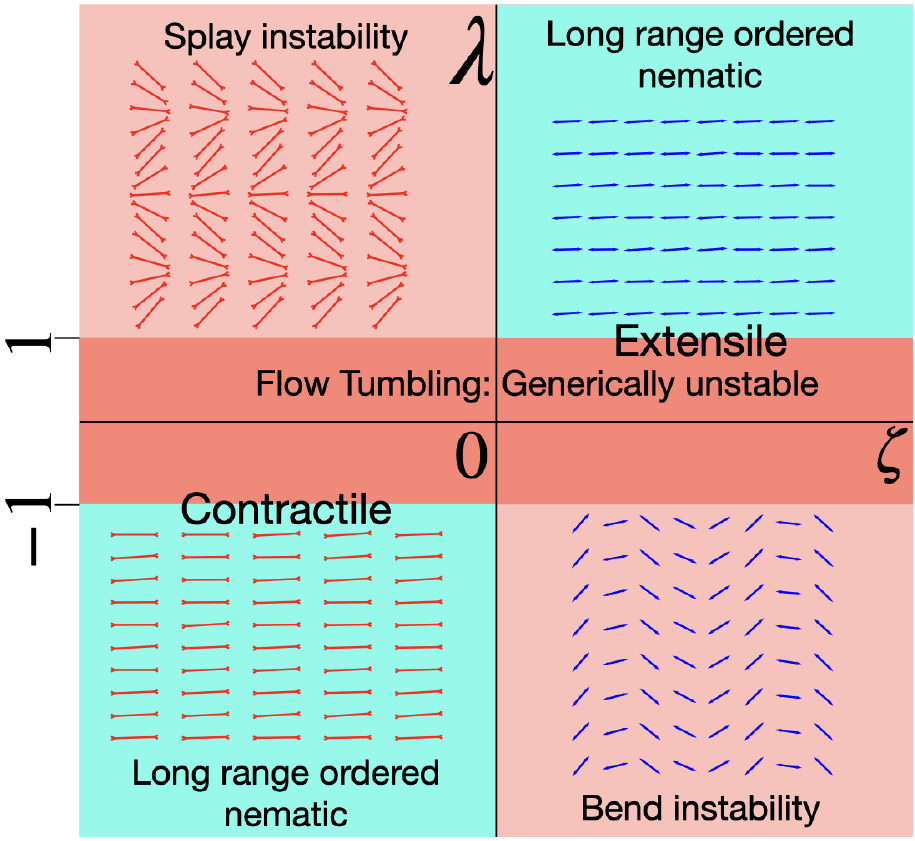}
	\caption{Stability diagram of active uniaxial phases at the $z=0$ interfacial plane between two fluids or at the boundary of a bulk fluid. $\zeta$ is the coefficient of active stress and $\lambda$ is the flow alignment parameter. For $|\lambda|>1$ and $\zeta\lambda>0$, a long-range ordered uniaxial phase is realised. Either when $\lambda\zeta<0$ or for $|\lambda|<1$ the ordered phase is destabilised for splay in contractile ($\zeta<0$) systems or for bend perturbations in extensile ($\zeta>0$) systems.}
	\label{Fig_phase}
\end{figure}
I expand \eqref{QtendynMT} and the active stress $\propto {\bsf Q}$ to linear order in angular fluctuations $\theta$ about an ordered state with $S=S_0=1$ (fluctuations of $S$ are non-hydrodynamic; see Supp. \ref{App252}). This yields
\begin{equation}
	\label{angdynstMT}
	\partial_t\theta=\Omega_{xy}-\lambda A_{xy}+\Gamma_\theta K\nabla_\perp^2\theta+\xi,
\end{equation}
where $\Gamma_\theta=\Gamma/4$ and $\xi$ is a non-conserving noise with variance $\Delta^Q/4$, from \eqref{QtendynMT}. The linear active force density is
${\bf f}^s=-i\zeta(q_y\theta\hat{x}+q_x\theta\hat{y})$. This in conjunction with \eqref{mobil_expMT} yields, to the lowest order in wavenumbers,
\begin{equation}
	\label{angfinMT}
	\partial_t\theta=-\frac{\zeta}{4\eta}\frac{q_x^4(\lambda-1)+q_y^4(\lambda+1)}{|q_\perp|^3}\theta+\mathcal{O}(q_\perp^2)+\xi\\=\frac{\zeta|q_\perp|}{4\eta}\left[\cos(2\phi)[1-\lambda\cos(2\phi)]-\frac{\lambda}{2}\sin^2(2\phi)\right]\theta+\mathcal{O}(q_\perp^2)+\xi,
\end{equation} 
where $\phi$ is the angle between ${\bf q}_\perp$ and the mean ordering direction $\hat{x}$.
Eq. \eqref{angfinMT} implies that the relaxation rate for angular fluctuations is \emph{positive} for all $\phi$ when $|\lambda|>1$ and $\zeta\lambda>0$ (see Supp. \ref{App252} \AM{for a detailed demonstration}). 
This directly demonstrates that a two-dimensional planar nematic phase is realised in the fully momentum-conserved NNC model (see Fig. \ref{Fig_phase}). 

\AM{Two-dimensional active nematic phases, both in the presence \cite{Ano_apol} and absence of fluids \cite{Aditi2, Chate1,Suraj_RG, Mishra_Chate} were thought to be possible \emph{only} in contact with substrates, which act as momentum sinks. While nematic suspensions in contact with a substrate \emph{can} retain order at arbitrarily high active drive \cite{Ano_apol}, that is a direct result of the momentum exchange with the substrate. This allows for an active force, that is not a divergence of a stress, with a distinct angular symmetry that tends to stabilise nematic ordering. In momentum-conserved systems, Simha-Ramaswamy instability was thought to forbid order. What allows interfacial states to evade this and order due to activity even though here the active force is a divergence of a stress? It was believed that one of the key requirements for the generic instability of aligned states is fluid incompressibility. The interfacial fluid velocity is \emph{not} incompressible \cite{Prasad, Boffeta}.} The interfacial flow \emph{only} becomes effectively incompressible when the surface is treated with surfactants \cite{com2, com3, com1} as in \cite{Guo1, Guo2, Sagues}. Moreover, all two-dimensional films in a three-dimensional medium, whether surfactant-coated or not, are compressible at large scales due to interfacial fluctuations \cite{Cai} even when they are incompressible at small scales. 
	Since I consider a clean interface, a non-vanishing dilational or compressive flow in the plane is compensated by a non-vanishing $z-$gradient of the three-dimensional velocity field (see  Eq. \eqref{bulkveltop} and \eqref{bulkvelbot} in Supp. \ref{App2} for a direct demonstration). Indeed, an in-plane director distortion leads to a compressive or dilational flow $i{\bf q}_\perp\cdot{\bf v}=(\zeta\theta/2\eta)(q_xq_y/|q_\perp|)=(\zeta\theta/4\eta)|q_\perp|\sin 2\phi$ that leads to the final term in the square brackets in \eqref{angfinMT} \AM{which is essential for
	ensuring the stability of nematic order.}

\AM{Is the compressibility of the interfacial flow \emph{sufficient} to ensure the stability of an ordered phase?
	Notice that the NNC model differs in two distinct ways from the conditions under which Simha-Ramaswamy instability is generally discussed: i. the interfacial fluid layer is compressible and ii. the number of active units at the interface is not conserved. To check whether the former by itself ensures the stability of the ordered phase, I will examine whether an ordered state in which the active units are constrained to live at the interface -- such that the number of particles at the interface is conserved (i.e. ii above doesn't hold) -- remains stable, in the next section.}

\AM{Now I examine the properties of the interfacial ordered state.} Since \eqref{angfinMT} implies that
the relaxation rate of angular fluctuations scales as $|q_\perp|$ along \emph{all} directions of the wavevector space -- unlike the relaxation rate of \AMF{equilibrium} nematics or active nematics on substrates both of which scale as $\sim q_\perp^2$ \cite{Aditi2, Ano_apol, RMP} -- the static structure factor of angular fluctuations $\langle|\theta({\bf q}_\perp,t)|^2\rangle\propto \Delta/|q_\perp|$.
Using this to calculate the depression of the order parameter from its perfectly ordered value $S_0$ due to fluctuations, I get $\langle S\rangle/ S_0=\langle \cos2\theta\rangle\equiv e^{-W}$ with $W=2\langle\theta({\bf r}_\perp,t)^2\rangle=2\int (d^2{q}_\perp/4\pi^2)\langle|\theta({\bf q}_\perp,t)|^2\rangle \propto\Lambda\Delta$, 
where $\Lambda$ is a wavenumber cut-off. 
Since $W$ remains finite, $\langle S\rangle$ does not vanish due to fluctuations even in infinite systems (for small enough $\Delta$). Thus, the linear theory predicts that an interfacial active nematic phase displays \emph{long-range order} (see Supp. \ref{App252} \AM{for an expanded version of this discussion}). \AM{The LRO phase here is a result of \emph{long-range interactions} due to the bulk fluid. If this interaction is cut-off at some (large) scale, for example, if instead of an interface in a bulk medium, one considers an interface in a thick layer of fluid resting on a substrate which acts as a momentum sink \cite{Alert}, then the order is quasi-long-ranged instead of long-ranged -- just as in active nematics in contact with a substrate \cite{Aditi2, Chate1,Suraj_RG, Mishra_Chate, Mahault, Ano_apol} -- as I show in Supp. \ref{App6}. Importantly, the ordered state is \emph{not} destabilised i.e., the mechanism via which the aligned state is stabilised does not require the fluid interaction to be long-ranged.}

To see whether \AM{the} conclusion \AM{about the interfacial flock having LRO} holds upon including the effect of nonlinearities, I use the standard renormalisation group logic: I rescale lengths, time and the angle field as $x\to b x$, $y\to b^\mu y$, $t\to b^z t$ and $\theta\to b^\chi\theta$, where $\mu$ is the anisotropy exponent, $z$ is the dynamical exponent and $\chi$ is the roughness exponent and examine whether any nonlinearity grows under this rescaling. Since $\partial_t\theta\sim -|q_\perp|\theta$ for all directions of the wavevector space, $z=\zeta=1$, within the linear theory. Further, since $\langle|\theta({\bf q}_\perp,t)|^2\rangle\sim \Delta/|q_\perp|$, $\chi=-1/2$. I now use these exponents to check the relevance of nonlinearities.
The most relevant nonlinearities that appear in \eqref{angfinMT} are from terms that have one power of the velocity field ${\bf v}$ and one power of $\theta$, along with a gradient operator, such as the one due to advection ${\bf v}\cdot\nabla_\perp\theta$. This scales as $q_\perp(\theta^2)_q$ since ${\bf v}\sim \bm{\Phi}(\phi)\theta$ where $\bm{\Phi}(\phi)$ is a vector function that depends on $\phi$, but not on $|q_\perp|$. This scales as $b^{z-1+\chi}=b^\chi$ \AM{which is} irrelevant at the Gaussian fixed point \AM{since $\chi<0$}. All other nonlinearites are \emph{even more} irrelevant implying that the linear exponents describe the \emph{exact} hydrodynamic properties of the LRO active nematic phase in a momentum-conserved system. 

Here, I demonstrated that the interfacial nematic phase is stable to \emph{spin-wave} fluctuations. In Supp. \ref{App26} I also argue that it should also be stable against unbinding of topological defects, confirming a 2d LRO nematic phase in 3d active fluids. In Supp. \ref{App27} I further demonstrate that polar flocks in the PNC model also display LRO with the \emph{same} exponents as the ones obtained here.

\section{Ordering when active units are confined to an interface}
\label{instabnem}
\AM{While the last section demonstrated that active nematic and polar boundary layers exist at interfaces in bulk fluids, the minimal requirements for escaping the Simha-Ramaswamy instability weren't clear. Are ordered phases possible in \emph{any} compressible system including momentum-conserved ones? To examine this, I consider systems in which active units are constrained to live only at the interface. The interfacial fluid velocity is still compressible, but the number of active particles at the interface is conserved. I first consider nematic ordering, and next, polar order.}
\AM{\subsection{Generic instability of a nematic phase in the NC model}}
\begin{figure}[!htb]
	\centering
	\includegraphics[width=6cm]{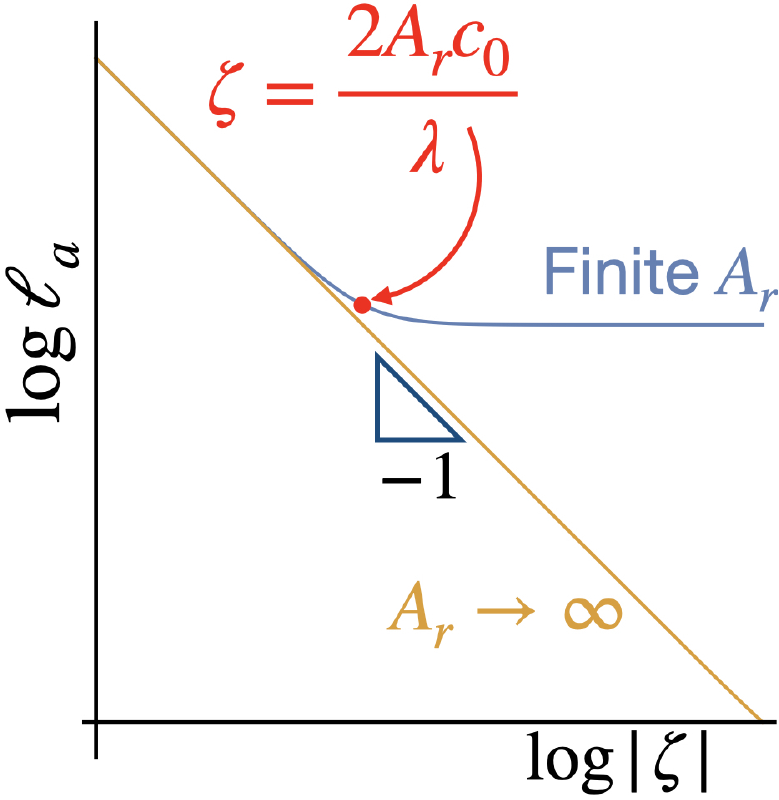}
	\caption{A log-log plot of the activity length scale $\ell_a$ -- the lengthscale associated with the fastest growing mode in active uniaxial fluids living at the interface between two fluids or at the boundary of a bulk medium-- as a function of activity $\zeta$. For an essentially incompressible system, i.e. $A_r\to\infty$, $\ell_a$ scale as $1/\zeta$ but for any finite $A_r$, $\ell_a$ scales as $1/\zeta$ only till $\zeta\sim 2 A_rc_0/\lambda$ saturating beyond that. %
	}
	\label{Fig_len}
\end{figure}
\AM{I now show that when active units are constrained to live at the interface, active nematic order is impossible i.e., nematic order is generically unstable in the NC model.}
To see this, I expand the equations of motion for concentration and angular fluctuations about a homogeneous, perfectly ordered state with $S_0=1$ and a mean concentration $c_0$. The linear fluctuations of the osmotic pressure about $c_0$ is taken to be $\Pi_c(c)\approx A_r(c_0)\delta c$ and the coupled linearised angular and concentration equations are
\begin{equation}
	\label{angfincMT}
	\partial_t\theta=-\frac{\zeta}{4\eta}\frac{q_x^4(\lambda-1)+q_y^4(\lambda+1)}{|q_\perp|^3}\theta-\frac{q_xq_yA_r\lambda}{4\eta|q_\perp|}\delta c+\xi,
\end{equation} 
and
\begin{equation}
	\label{conceqfinMT}
	\partial_t\delta c=-\frac{\zeta c_0 q_xq_y}{2\eta|q_\perp|}\theta-\frac{c_0|q_\perp|A_r}{4\eta}\delta c.
\end{equation}
Here all coefficients such as $\zeta$, $\lambda$ are evaluated at $c_0$ (this is a simplified limit of the full coupled equations of motion for $\delta c$ and $\theta$ displayed in Supp. \ref{App45}\AMF{; here I assume that $\zeta$ is not a function of $c$}). 
One of the two eigenfrequencies implied by \eqref{angfincMT} and \eqref{conceqfinMT} changes sign as the angle between the wavevector of perturbation and the ordering direction, $\phi$ passes through \AM{$\pi/4$}:
	\begin{equation}
		\label{asympom3MT}
		\omega_-(\phi\approx\pi/4)\approx-\frac{i|q_\perp|}{\eta}\frac{A_rc_0\zeta}{2A_rc_0+\zeta\lambda}\left[\phi-\frac{\pi}{4}\right],
	\end{equation}
implying an instability of the ordered state for perturbations with $\phi$ either just above or just below $\pi/4$.

	Interestingly, this eigenfrequency becomes \emph{independent} of $\zeta$, and therefore, of its sign (i.e., its contractile or extensile character), at small $A_r$ (i.e., when $\zeta\lambda\gg A_r c_0$. Of course, $\omega_-$ vanishes when $\zeta=0$). 
	Therefore, $\omega_-$ crosses over from being independent of $\zeta$ when $\zeta\lambda\gg A_r c_0$ to linearly depending on $\zeta$ when $\zeta\lambda\ll A_r c_0$. That the character of the instability can be controlled by changing the concentration of the active particles (which controls $A_r$), has important consequences for the experiments on motor-microtubule films \cite{Sagues, Alert}. Often, these experiments measure the ``activity lengthscale'' $\ell_a$. This scale is associated with the fastest growing mode obtained by retaining $\mathcal{O}(q_\perp^2)$ terms in \eqref{angfincMT} and \eqref{conceqfinMT} that stabilise the dynamics at larger wavenumbers \AM{(this explicitly depends on the Saffman-Delbr\"uck length; see Supp. \ref{App3})}. 
	The form of \eqref{asympom3MT} demonstrates that for small enough $A_r$, $\ell_a \propto {(2A_rc_0+\zeta\lambda)}/{A_rc_0\zeta}$ saturates at a finite value as activity is increased, while for larger $A_r$, it scales as $1/\zeta$ (see Fig. \ref{Fig_len}). This is in contrast to current predictions that $\ell_a$ decreases monotonically with activity \cite{Sagues, Alert}. 
	This implies that the fastest growing mode, or the typical vortex size in the spatiotemporally chaotic state, in motor-microtubule experiments \cite{Sagues} should become independent of activity as the density of the active particles is reduced; i.e., at low microtubule concentration, $\ell_a$ should become independent of the kinesin concentration.

	 \AM{The generic instability of the nematic state in this model implies that the lack of two-dimensional incompressibility of an interfacial layer is not enough to stabilise nematic order. Instead, it also requires the number of active units at the interface to not be conserved. Indeed, while Simha-Ramaswamy instability has, till now, been thought to be associated with \emph{incompressible} bulk fluids, I show in Supp. \ref{App7} that bulk nematic order is unstable -- with the instability having a character similar to the one discussed in this section (albeit with a wavenumber-independent growth rate) -- in any momentum and mass-conserved fluid \AMF{and not just an incompressible one}. This provides a rationalisation for nematic interfacial order additionally requiring an exchange of active particles between the interface and the bulk.}

\AM{\subsection{Stable polar phase PC model: Outrunning the generic instability}}
\label{stabpol}
\begin{figure}[!htb]
	\centering
	\includegraphics[width=7cm]{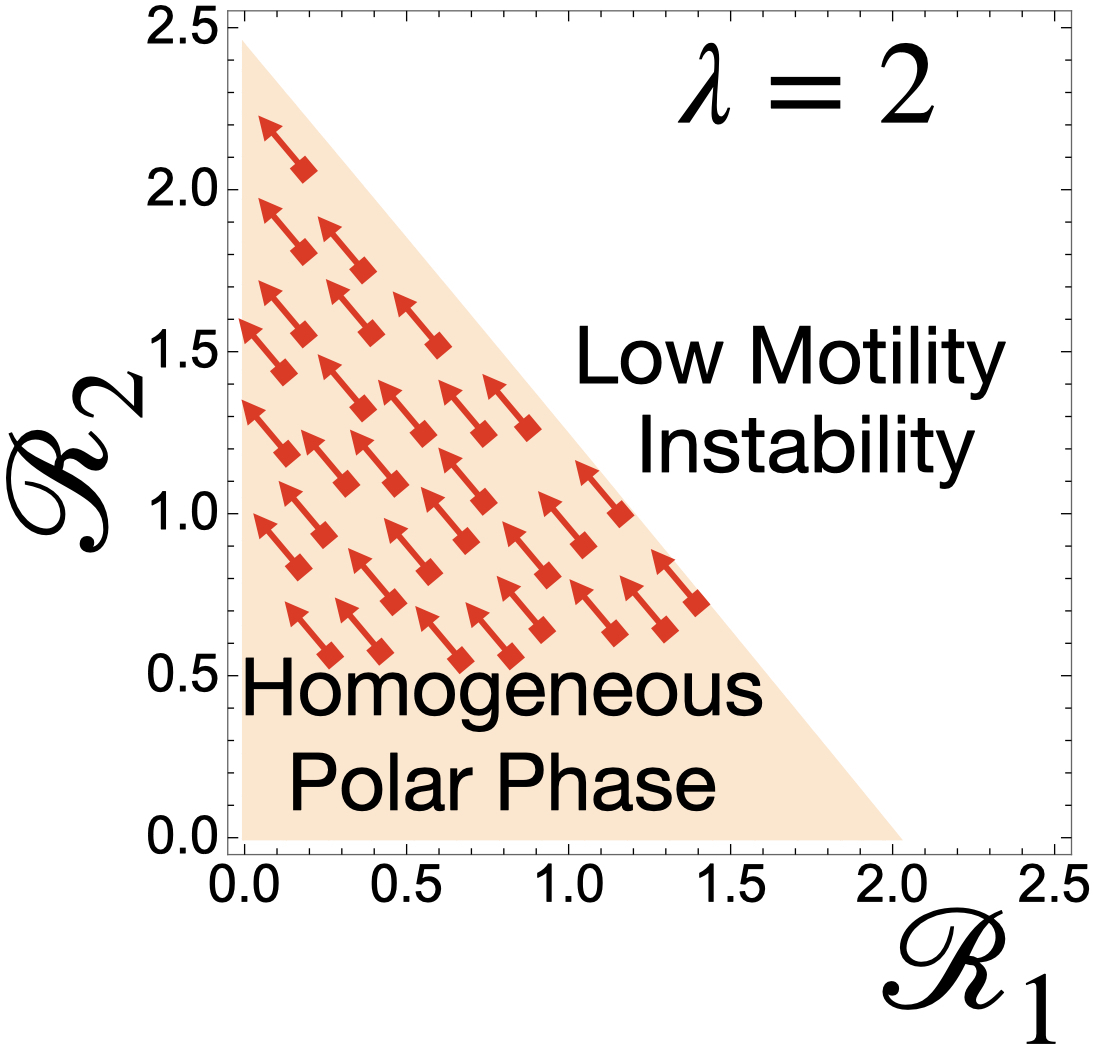}
	\caption{Stability region for a motile flock living either at a two-fluid interface or at the boundary of a bulk fluid in the $\mathscr{R}_1-\mathscr{R}_2$ plane for a specific value of $\lambda=2$ implied by \eqref{orunfullMT}. $\mathscr{R}_1$ and $\mathscr{R}_2$ are proportional to the ratio of the inverse compressibility $A_r$ to motility $v_p$ and the strength of active stress $\zeta$ to motility respectively.
		The stability region increases at higher $\lambda$.}
	\label{Fig_stab}
\end{figure}
While apolar active particles constrained to live at surfaces of momentum-conserving fluids cannot order, surprisingly, \emph{polar} particles can. \AM{That is, while both bulk polar and apolar order are generically unstable in Stokesian, momentum-conserved active fluids (see Supp. \ref{App7}), the minimal conditions for the existence of an \emph{interfacial }ordered state differ for the two kinds of order. Apolar order \emph{only exists} in the absence of any interfacial conserved quantity but polar order is more robust.} Highly motile units outrun the instability suffered by their immotile counterparts. Here I demonstrate this in a particularly simple case (see Supp. \ref{App46} for the discussion of the full model) where the \emph{only} polar term I retain is the self-advection of the angular fluctuations. That is, the concentration dynamics is still described by \eqref{conceqfinMT} (i.e., I artificially set the active motility to $0$ for this illustration; see Supp. \ref{App46}) and I add a term $-iv_pq_x\theta$ to the R.H.S. of \eqref{angfincMT}. This turns out to be enough to ensure the existence of a stable active polar phase.
The eigenfrequencies, for fluctuations about a polarised state, written in terms of two dimensionless numbers
$\mathscr{R}_1=A_rc_0/4v_p\eta$ and $\mathscr{R}_2=\zeta/4v_p\eta$ are
\begin{widetext}
	\begin{equation}
		\label{orunfullMT}
		\omega_\pm=\frac{i|q_\perp|v_p}{2}\left[\mathcal{S}(\phi)\pm\sqrt{\mathcal{S}(\phi)^2-4\mathscr{R}_1\{i\cos\phi-\mathscr{R}_2\cos2\phi(1-\lambda\cos2\phi)\}}\right],
	\end{equation}
\end{widetext}
where $\mathcal{S}(\phi)=-\mathscr{R}_1+\mathscr{R}_2[\cos2\phi-\lambda(1+\cos^2\phi)/2]-i\cos\phi$. 
The stability boundary of the polar phase depends on the non-dimensional parameters $\mathscr{R}_1$, $\mathscr{R}_2$ and $\lambda$. A representative plot of the stability region of a homogeneous flock in the $\mathscr{R}_1$, $\mathscr{R}_2$ plane, for a specific choice of $\lambda$ ($\lambda=2$), is displayed in Fig. \ref{Fig_stab}.

I show in Supp. \ref{App46} that the linear theory describes the \emph{exact} hydrodynamic behaviour of this flock and the fluctuations have the same exponents that characterise nematic and polar flocks in the NNC and PNC models. Further, considering the full PC model (i.e., one in which the coefficient of the active \AM{motility} has not been set to $0$), I find (see Supp. \ref{App46}) that the roughness exponent for the concentration fluctuations, $\chi_c=-1/2$ as well. This implies that the equal-time concentration fluctuations $\langle|\delta c({\bf q}_\perp,t)|^2\rangle\sim 1/|q_\perp|$ in the polar phase. The divergence at small wavenumbers implies that R.M.S. number fluctuations $\sqrt{\langle \delta N^2\rangle}$ in a region {with $\langle  N\rangle$ particles on average scales} as $\langle N\rangle^{3/4}$, instead of as $\langle N\rangle^{1/2}$ as it would in all equilibrium systems not at a critical point, and thus violates the law of large numbers. Such \emph{giant number fluctuations} -- albeit with a different exponent -- had earlier been discussed for active systems \emph{in contact with substrates} \cite{RMP, Aditi2, Toner_rean, SRrev, Ano_apol} but here appears in a fully-momentum-conserved system. 

\AM{The mechanism via which polar order is stabilised even when the active units are constrained to live at the interface bears some resemblance to how motility stabilises \emph{bulk} flocks in inertial fluids \cite{Rayan}. However, here the interfacial flock is stabilised even in the strict Stokesian regime.}
	{

\section{Conclusions}
\label{conc}
In this article, I have shown that active orientable particles form ordered phases at interfaces of momentum-conserved bulk fluids. \AM{While nematic order requires the exchange of particles between the bulk and the interface, polar order does not.}
In appendix \ref{App5} I further demonstrate that small fluctuations of the interface itself do not destroy this ordering. This is particularly relevant for order on membranes immersed in a bulk fluid.

I now discuss some experimental systems and realistic situations in which the results obtained in this article should be observed and tested. 
Self-assembled motor-microtubule layers at oil-water interfaces \cite{Dogic1, Dogic2, Sagues} or in bulk fluids \cite{Ano_und, Ramin_und} are one of the mainstays of studying active pattern formation. In current experiments, the number of capped microtubule filaments at the interface is essentially constant and therefore, the nematic state is generically unstable, as expected. Further, due to the surfactants present in these systems \cite{Lemma, Guillamat}, the interfacial layer is essentially incompressible in two dimensions. %
However, the incompressibility constraint may be removed and the value of $A_r$ in these experiments may be varied by not using surfactants and changing concentrations of microtubule filaments. As $A_r$ is varied, the activity lengthscale --  or the scale of the vortices in the chaotic state -- \AM{should display} the behaviour described in  Sec. \ref{instabnem}. More radically, an ordered nematic phase may be realised by allowing the microtubule filaments to associate and dissociate at the interface and the monomers to diffuse in the fluid. 
A further complication is that these experiments \cite{Dogic1, Dogic2, Sagues, Alert,Ano_und, Ramin_und} are performed in confined channels. This cuts off the long-range fluid interactions, \AM{but} I demonstrate in appendix \ref{App6}, that a stable uniaxial phase is still realised at arbitrarily high values of activity. 
However, the suppression of long-range interactions \AM{enhances orientational fluctuations leading to only} quasi-long-range order.

Beyond experiments on intracellular gels, the theory presented here naturally models boundary layers of bacteria that aggregate at air-water interfaces \cite{bac_int1, bac_int2, bac_int3, bac_int4}. Such bacterial films may display ordered nematic phases predicted in Sec. \ref{stabNNC}. Interfacial \emph{active} nematics can also be created by impregnating an \AMF{equilibrium} interfacial nematic with bacteria. 

Active polar phases are often associated with membranes \cite{Ano_mem} in cellular systems and have important consequences for signalling and transport. One particularly interesting example of this is the polar ordering of short actomyosin filaments at cell membranes that have been argued to be crucial for nanoclustering of cell-surface molecules \cite{jitu1, jitu2, Madan1, Madan2}. Cell membranes are in contact with \AM{active} bulk cytoskeletal fluid 
\AM{which may} promote ordering and, thus, membrane-associated transport.

A route to the spontaneous generation of surfaces is via phase separation which is common in active systems \cite{modelH2} and this article demonstrates that such surfaces can naturally host a flocking wetting layer.
 In active systems, small droplets can form due to the segregation of molecules in a complex mixture by phase separation \cite{Jul1}. These droplets can grow and separate due to chemical activity thus forming protocells. If some of the components in these protocells are elongated, they may form a stable and aligned boundary layer at the edge of the protocell leading to an aligned protocortex. 
 
 I look forward to quantitative and qualitative experimental examinations of properties of ordered active wetting layers.

\begin{acknowledgments}
	I thank Sriram Ramaswamy and Raphael Voituriez for insightful comments and discussions. I also thank Sriram Ramaswamy for his careful reading of the manuscript and for crucial suggestions. 
	I thank a Talent fellowship awarded by the CY Cergy Paris University.
\end{acknowledgments}

\end{document}